%% file: ifacconf.tex
\documentclass{ifacconf}

\usepackage{graphicx}      %
\usepackage{natbib}        %
\usepackage{amsmath}
\usepackage{amssymb}
\usepackage{mathrsfs}
\usepackage{algorithm}
\usepackage[noend]{algpseudocode}
\usepackage{caption}
\usepackage{subcaption}
\usepackage{tikz}
\usepackage{pgfplots}
\usetikzlibrary{backgrounds,intersections}

\usetikzlibrary{external}
\DeclareMathOperator{\tr}{tr}
\begin{document}
\begin{frontmatter}

\title{Safe and Efficient Switching Controller Design for Partially Observed Linear-Gaussian Systems}

\author[First]{Yiwen Lu} 
\author[First]{Yilin Mo} 

\address[First]{Department of Automation and BNRist, Tsinghua University, Beijing, China (e-mail: luyw20@mails.tsinghua.edu.cn, ylmo@tsinghua.edu.cn).}

\begin{abstract}                %
    Switching control strategies that unite a potentially high-performance but uncertified controller and a stabilizing albeit conservative controller are shown to be able to balance safety with efficiency, but have been less studied under partial observation of state. To address this gap, we propose a switching control strategy for partially observed linear-Gaussian systems with provable performance guarantees. We show that the proposed switching strategy is both safe and efficient, in the sense that: (1) the linear-quadratic cost of the system is always bounded even if the original uncertified controller is destabilizing; (2) in the case when the uncertified controller is stabilizing, the performance loss induced by the conservativeness of switching converges super-exponentially to zero. The effectiveness of the switching strategy is also demonstrated via numerical simulation on the Tennessee Eastman Process.
\end{abstract}


\end{frontmatter}

\input{intro.tex}
\input{problem.tex}
\input{theory.tex}

\input{simulation.tex}

\input{conclusion.tex}

\bibliography{ref}             %

\ifdefined\excludeappendix
\else
    \appendix
    \input{appendix.tex}

\fi

\end{document}

%% file: intro.tex
\section{Introduction}

\begin{figure}[!htbp]
    \centering
    \includegraphics{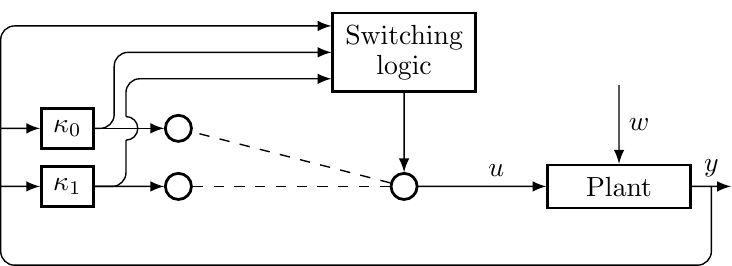}
    \caption{Illustration of the switching strategy under discussion. The blocks $\kappa_1, \kappa_0$ stand for primary and fallback controllers respectively. The switching logic chooses whether the control input determined by $\kappa_1$ or $\kappa_0$ is applied to the plant.}
    \label{fig:block_diagram}
\end{figure}

A class of switching strategies has recently been designed to control a system by uniting a primary controller, which is potentially high-performance but uncertified, and a fallback controller, which is guaranteed to be stabilizing but typically conservative ~\citep{cdc,wintz2022global,wang2021exact}. These strategies can be illustrated by Fig.~\ref{fig:block_diagram}. It is desirable that the primary controller is applied most of the time during normal operation, but the switching strategy provides an additional layer of safeguard by falling back to a stabilizing controller on the detection of safety breach, characterized by large system states.
This can be particularly useful in the context of adaptive and data-driven control, where it is often difficult to guarantee the stability of closed-loop systems under controllers that are learned or designed using identified system models.

In this paper, we propose and analyze a switching control strategy for \emph{partially observed} linear systems driven by Gaussian noise following the above principle.
The aforementioned existing works make the strong assumption that the system state can be measured directly and the controllers are \emph{static} state feedback controllers.
By contrast, in the setting considered in this paper, the system state can only be inferred from noisy output, and both the primary and fallback controllers are \emph{dynamic}, i.e., they may maintain internal states. This partially observed setting complicates the analysis by introducing the interplay between system and controller states, but is crucial if the control strategy is to be deployed to real systems under limited measurement capabilities.

The proposed switching control strategy is analyzed from both stability and optimality aspects. In particular, we prove the following results:
\begin{enumerate}
    \item The LQ cost under the proposed strategy is always \emph{bounded} for \emph{any} primary controller, which implies that the proposed strategy improves the safety of the uncertified primary controller by preventing the system from being catastrophically destabilized.
    \item Provided that the primary controller is stabilizing, the additional LQ cost compared to using the primary controller alone scales as $\mathcal{O}(t^{1/4}\exp(-\text{constant}\cdot M^2))$, where the dwell time $t$ and the switching threshold $M$ are tunable parameters of the strategy. The super-exponential decay of the above quantity as the threshold $M$ increases implies the proposed strategy is efficient in the sense that the induced performance loss quickly becomes negligible as the strategy is tuned to be less conservative.
\end{enumerate}
The above theoretical results are validated by simulation on the Tennessee Eastman Process example.

\subsection*{Related works}

\paragraph*{Switched control systems}

Switching controllers known as supervisors have long been used to stabilize certain classes of nonlinear systems~\citep{hespanha1996supervision,teel1997uniting,seto1998simplex,prieur2001uniting,el2005output,battistelli2012supervisory}. Most of the above works focus on the stability of closed-loop systems, while the (near-)optimality of the controllers are less discussed.
\cite{wintz2022global} propose the idea of switching between certified and uncertified controllers to improve the control performance for nonlinear state feedback systems, but does not provide a quantitative analysis of the performance under switching.
\cite{cdc} quantify the worst-case cost and performance loss under switching for linear state feedback system, and \cite{wang2021exact} adopts a similar switching strategy for closed-loop identification and adaptive control. However, the above two works assume full observation of the state, which may be impractical.
To our knowledge, the quantitative analysis of switching controller for partially observed linear systems has not been widely studied.

\paragraph*{Adaptive control of partially observed linear systems}

The data-driven and learning-based control of partially observed linear systems, especially the adaptive Linear-Quadratic-Gaussian (LQG) problem, has drawn significant research attention in recent years, in both offline~\citep{mania2019certainty,tsiamis2020sample,zheng2021sample} and online~\citep{lale2020logarithmic,simchowitz2020improper,ziemann2022regret} settings.
These works usually provide high-probability performance bounds rather than strict convergence guarantees, partially due to the always nonzero probability of the system being destabilized by a controller learned from finite noise-corrupted data.
Since the switching control strategy proposed in the current manuscript is agnostic to how the primary controller is obtained, it may potentially serve as a ``plug-and-play'' modification to the existing adaptive LQG algorithms that enhances the safety of learned controllers.

\paragraph*{Nonlinear controller for linear systems} The study of nonlinear controllers for linear systems has mainly focused on engineering aspects such as saturating actuators. The performance of such controllers has been studied in \cite{gokcek2001lqr} using stochastic linearization, a heuristics for replacing nonlinearity with approximately equivalent gain and bias.
By contrast, in the current paper, by introducing a switching control strategy, the nonlinearity of controller is a design choice rather than a physical constraint, and rigorous performance bounds are derived without resorting to heuristics.

\subsection*{Outline}

The remainder of this manuscript is organized as follows: Section~\ref{sec:formulation} describes the problem setting and formalizes the proposed switching control strategy. The properties of the proposed strategy are derived in Section~\ref{sec:thoery}. Section~\ref{sec:simulation} verifies the theoretical results using simulation on an industrial process example. Finally, Section~\ref{sec:conclusion} summarizes the manuscript.

\subsection*{Notations}

The set of real numbers and the set of nonnegative integers are denoted by $\mathbb{R}, \mathbb{N}$ respectively.
For a matrix $M$, the transpose of $M$ is denoted by $M^\top$.
For a square matrix $M$, the spectral radius of $M$ is denoted by $\rho(M)$, and the trace of $M$ is denoted by $\tr(M)$.
For a real symmetric matrix $M$, we say $M \succ 0$ if $M$ is positive definite.
For a vector $v$, the 2-norm of $v$ is denoted by $\| v \|$, and for a matrix $M$, the induced 2-norm of $M$ is denoted by $\| M \|$.
For $P \succ 0$, the $P$-norm of a vector $v$ with the proper dimension is defined by $\| v \|_P = \| P^{1/2} v\|$, and the induced $P$-norm of a square matrix $M$ with proper dimensions is defined by $\| M \|_P = \| P^{-1/2} M P^{-1/2} \|$.
A Gaussian random vector $X$ with mean $\mu$ and covariance $\Sigma$ is denoted by $X \sim \mathcal{N}(\mu, \Sigma)$.
The probability and expectation operators are denoted by $\mathbb{P}(\cdot), \mathbb{E}(\cdot)$ respectively, and $\mathbf{1}_E$ is the indicator function of a random event $E$.
For functions $f(x),g(x)$ with nonnegative values, we say $f(x) = \mathcal{O}(g(x))$ if $\limsup_{x\to\infty} f(x)/g(x) < \infty$.

%% file: problem.tex
\section{Problem Formulation and Proposed Control Strategy}
\label{sec:formulation}

Consider the discrete-time Linear-Quadratic-Gaussian\\(LQG) control setting: the system is
\begin{equation}
    \begin{cases}
        x(k+1) = Ax(k) + Bu(k) + w(k), \\
        y(k) = Cx(k) + v(k),
    \end{cases}
\end{equation}
where the time index is denoted by $k\in \mathbb{N}$, the state, input and output vectors are denoted by $x(k) \in \mathbb{R}^n, u(k) \in \mathbb{R}^m, y(k) \in \mathbb{R}^p$ respectively, and the process and measurement noise vectors are denoted by $$w(k) \stackrel{\text { i.i.d. }}{\sim} \mathcal{N}(0, W), v(k) \stackrel{\text { i.i.d. }}{\sim} \mathcal{N}(0, V)$$ respectively, where $W \succ 0, V \succ 0$.
We assume w.l.o.g. that $(A, B)$ is controllable and $(A,C)$ is observable.
The performance of a controller is measured in terms of the infinite-horizon LQ cost defined as:
\begin{equation}
    J=\limsup _{T \rightarrow \infty} \frac{1}{T} \mathbb{E}\left[\sum_{k=0}^{T-1} x(k)^{\top} Q x(k) +u(k)^\top R u(k)\right],
    \label{eq:J}
\end{equation}
where $Q \succ 0, R \succ 0$ are fixed weight matrices. It is well known that a controller for the above system can be designed by combining a Luenberger observer with a feedback controller. In particular, when $K, L$ are matrices of proper dimensions such that both $A+BK$ and $A-LC$ are stable, a stabilizing controller can be specified as follows:
\begin{equation}
    \begin{cases}
        \hat{x}(k+1) = (A - LC) \hat{x}(k) + B u(k) + L y(k), \\
        u(k) = K \hat{x}(k),
    \end{cases}
    \label{eq:oracle}
\end{equation}
where $\hat{x}(k) \in \mathbb{R}^n$ is an estimate of the true state. The controller is optimal when the matrices $K, L$ are the optimal feedback gain and the Kalman gain respectively, both of which can be determined by solving discrete-time algebraic Riccati equations.

When the system model $(A, B, C, W, V)$ is unknown, however, the optimal controller cannot be computed directly. Instead, we assume the availability of the following two controllers, both specified in the general form of linear dynamic controllers:
\begin{itemize}
    \item Primary controller: \begin{equation}
        \begin{cases}
            z_1(k+1) = A_1 z_1(k) + B_1 u(k) + L_1 y(k), \\
            u_1(k) = K_1 z_1(k),
        \end{cases}
        \label{eq:primary}
    \end{equation}
    with internal state $z_1(k) \in \mathbb{R}^{n_1}$. This controller is typically learned from data, which may achieve near-optimal performance, but does not have stability guarantees.

    \item Fallback controller: \begin{equation}
        \begin{cases}
            z_0(k+1) = A_0 z_0(k) + B_{0} u(k) + L_0 y(k), \\
            u_0(k) = K_0 z_0(k),
        \end{cases}
        \label{eq:backup}
    \end{equation}
    with internal state $z_0(k) \in \mathbb{R}^{n_0}$, which is typically conservative but always guaranteed to be stabilizing.
    The existence of such a known stabilizing controller is commonly assumed in the literature on LQ adaptive control~\cite{wang2021exact,lu2023almost,ziemann2022regret}.
    In particular, we make the following assumption:

    \begin{assum}
        The matrix
        \begin{equation}
            \mathcal{A}_0 := \begin{bmatrix}
                A & B K_0 \\
                L_0 C & A_0 + B_0 K_0
            \end{bmatrix}
            \label{eq:A0}
        \end{equation}
        is stable, and the matrix $A_0$ is also stable.
        \label{assum:A0}
    \end{assum}

    \begin{rem}
        The matrix $\mathcal{A}_0$ defined in~\eqref{eq:A0} is the system matrix of $[x^\top \; z_0^\top]^\top$, which must be stable if the controller is to stabilize both the system state and its internal state.

        Meanwhile, for the controller~\eqref{eq:oracle} designed with the known system model, it holds $A_0 = A-LC$, which is stable for a properly designed Luenberger observer. Therefore, it is reasonable to assume that $A_0$ is also stable for the fallback controller. Furthermore, whether $A_0$ is stable can be easily verified since it is a known controller parameter.
    \end{rem}
    
\end{itemize}

A supervisor can be deployed to select the actual input $u(k)$ from the candidate inputs $u_1(k), u_0(k)$ specified by the primary and fallback controllers respectively. Ideally, one may desire to use $u_1(k)$ as often as possible since it usually admits a better performance, but switch to $u_0(k)$ when potential instability is detected. Although the system state cannot be directly measured, a large difference between the uncertified input \( u_1(k) \) and the stabilizing input \( u_0(k) \) may be indicative of instability.
Motivated by the above intuition, the proposed control strategy normally applies $u_1(k)$, while falling back to $u_0(k)$ for $t$ consecutive steps when $\left\| u_1(k) - u_0(k) \right\|$ exceeds a threshold $M$. The procedure is formally described in Algorithm~\ref{alg:main}.

\renewcommand{\algorithmicrequire}{\textbf{Input:}}
\renewcommand{\algorithmicensure}{\textbf{Initialize:}}
\begin{algorithm}[!htbp]
    \begin{algorithmic}[1]
        \Require Switching threshold $M$, dwell time $t$
        \Ensure $\xi(0) = 0, z_0(0) = 0, z_1(0) = 0$ \texttt{//$\xi(k) \in \mathbb{N}$ is an internal state of the supervisor which counts how many remaining steps to use the fallback controller}
        \For{$k = 0, 1, \ldots$}
            \State Compute $u_1(k), u_0(k)$ using~\eqref{eq:primary} and~\eqref{eq:backup}
            \If{$\xi(k) > 0$}
                \State $u(k) \gets u_0(k)$
            \Else
                \If{$ \| u_1(k) - u_0(k) \|  \geq M$}
                    \State $\xi(k) \gets t, u(k) \gets u_0(k)$
                \Else
                    \State $u(k) \gets u_1(k)$
                \EndIf
            \EndIf
            \State Apply $u(k)$ to system
            \State Update $z_1(k+1), z_0(k+1)$ using~\eqref{eq:primary} and~\eqref{eq:backup} \texttt{// Updates of both internal states use $u(k)$ instead of the respective $u_1(k), u_0(k)$}
            \State Update $\xi(k+1) \gets \max\{\xi(k) - 1, 0 \}$
        \EndFor
    \end{algorithmic}
    \caption{Proposed switching control strategy}
    \label{alg:main}
\end{algorithm}

The performance of the proposed control strategy can be evaluated in terms of: \begin{itemize}
    \item Safety: the LQ cost should be upper bounded, regardless of the choice of the primary controller;
    \item Efficiency: the increase in LQ cost caused by switching should be small when the primary controller is stabilizing.
\end{itemize}
Therefore, the next section is dedicated to analyzing the safety and efficiency of the proposed strategy.

%% file: theory.tex
\section{Theoretical Results}
\label{sec:thoery}

This section is devoted to proving the boundedness of the closed-loop system under the proposed switching strategy as well as quantifying performance loss that switching incurs.
Proofs of the results are omitted from the main text due to space limit, and readers are referred to the appendix of the online version of this paper~\cite{arxiv} for the proofs.

\subsection{Upper Bound on the LQ Cost}

By Assumption~\ref{assum:A0}, there exists $P_0 \succ 0$ which is the solution to the discrete-time Lyapunov equation
\begin{equation}
    \mathcal{A}_0^\top P_0 \mathcal{A}_0 - P_0 + \begin{bmatrix}
        Q & 0 \\
        0 & K_0^\top R K_0 + I
    \end{bmatrix} = 0,
    \label{eq:P0}
\end{equation}
and hence there exists $0 < \rho_0 < 1$ such that
\begin{equation}
    \mathcal{A}_0^\top P_0 \mathcal{A}_0 \prec \rho_0 P_0.
    \label{eq:rho0}
\end{equation}

An upper bound on the LQ cost can be derived via the following Lyapunov function:
\begin{equation}
    V_0(k) = \begin{bmatrix}
        x(k) \\
        z_0(k)
    \end{bmatrix}^\top P_0 \begin{bmatrix}
        x(k) \\
        z_0(k)
    \end{bmatrix}.
\end{equation}

\begin{lem}
    It holds for any $k$ that
    \begin{align}
        & \mathbb{E}V_0(k) \leq \frac{4(1+\rho_0)}{(1 - \rho_0)^2}\cdot \nonumber \\
        & \left(  M^2 \left\|\begin{bmatrix} B \\ B_0 \end{bmatrix} \right\|^2 \| P_0 \| + \tr \left( \begin{bmatrix}
            W & 0 \\
            0 & L_0 VL_0^\top
        \end{bmatrix} P_0 \right) \right),
    \end{align}
    where $P_0, \rho_0$ are defined in~\eqref{eq:P0} and~\eqref{eq:rho0}.
    \label{lem:V0}
\end{lem}

\begin{thm}
    For any controller parameters \( M,t \), the LQ cost under the proposed control strategy satisfies
    \begin{align}
        & J \leq \left( \frac{8(1+\rho_0)\left\| \begin{bmatrix} B \\ B_0 \end{bmatrix} \right\|^2 \| P_0 \|}{(1 - \rho_0)^2}+ 2\| R \| \right)M^2 + \nonumber \\
        & \frac{8(1+\rho_0)\tr \left( \begin{bmatrix}
            W & 0 \\
            0 & L_0VL_0^\top
        \end{bmatrix} P_0 \right)}{(1 - \rho_0)^2},
        \label{eq:safe}
    \end{align}
    where $P_0, \rho_0$ are defined in~\eqref{eq:P0} and~\eqref{eq:rho0}.
    \label{thm:safety}
\end{thm}

\subsection{Upper bound on performance loss caused by switching}

In this subsection, we quantify the extra LQ cost caused by the conservativeness of switching, under the following assumption on the stability of primary controller which parallels Assumption~\ref{assum:A0}:

\begin{assum}
    The matrix
    \begin{equation}
        \mathcal{A}_1 := \begin{bmatrix}
            A & B K_1 \\
            L_1 C & A_1 + B_1 K_1
        \end{bmatrix}
        \label{eq:A1}
    \end{equation}
    is stable, and the matrix $A_1$ is also stable.
    \label{assum:A1}
\end{assum}

The behavior of the closed-loop system under switching can be characterized via a common Lyapunov function. However, since both the primary and fallback controllers are \emph{dynamic}, Lyapunov function should be defined on all states of the closed-loop system, instead of the state of the open-loop system. In particular, consider the augmented system whose state $\mathscr{X}(k)$ is stacked from the state of the plant and the internal states of the two controllers, and whose noise $\mathscr{W}(k)$ is stacked from the process noise and the measurement noise magnified by $L_0$ and $L_1$ respectively: denote
\begin{equation}
    \mathscr{X}(k) = \begin{bmatrix}
        x(k) \\ z_0(k) \\ z_1(k)
    \end{bmatrix},
    \mathscr{W}(k) = \begin{bmatrix}
        w(k) \\ L_0 v(k) \\ L_1 v(k)
    \end{bmatrix},
\end{equation}
then the system matrices of $\{ \mathscr{X}(k) \}$ under the primary and fallback controllers respectively are:
\begin{equation}
    \mathscr{A}_1 := \begin{bmatrix}
        A & 0 & BK_1 \\
        L_0C & A_0 & B_1 K_1 \\
        L_1C & 0 & A_1 + B_1K_1
    \end{bmatrix},
    \label{eq:A_aug_1}
\end{equation}
and
\begin{equation}
    \mathscr{A}_0 := \begin{bmatrix}
        A & BK_0 & 0 \\
        L_0C & A_0 + B_0K_0 & 0 \\
        L_1C & B_0K_0 & A_1
    \end{bmatrix}.
    \label{eq:A_aug_0}
\end{equation}
Since $\mathscr{A}_0$ is block lower-diagonal with the diagonal blocks $\mathcal{A}_0, A_0$ being stable by Assumption~\ref{assum:A0}, the matrix $\mathscr{A}_0$ is stable. Similarly, by Assumption~\ref{assum:A1}, the matrix $\mathscr{A}_1$ is also stable.

The above defined augmented system evolves as
\begin{equation}
    \mathscr{X}(k+1) = \begin{cases}
        \mathscr{A}_1 \mathscr{X}(k) + \mathscr{W}(k) & u(k) = u_1(k), \\
        \mathscr{A}_0 \mathscr{X}(k) + \mathscr{W}(k) & u(k) = u_0(k),\\
    \end{cases}
\end{equation}
and $\mathscr{W}(k) \sim \mathcal{N}(0, \Sigma)$, where
\begin{equation}
    \Sigma := 
    \begin{bmatrix}
        W & 0 & 0 \\
        0 & L_0VL_0^\top & L_0 V L_1^\top \\
        0 & L_1 V L_0^\top & L_1 V L_1^\top
    \end{bmatrix}.
    \label{eq:Sigma}
\end{equation}

Since both $\mathscr{A}_0$ and $\mathscr{A}_1$ are stable, the following inequalities hold simultaneously for sufficiently large dwell time $t$:
\begin{equation}
    \begin{cases}
        \mathscr{A}_1^\top P \mathscr{A}_1 < \rho P, \\
        ( \mathscr{A}_0^t)^\top P \mathscr{A}_0^t < \rho P,
    \end{cases}
    \label{eq:common_lyap}
\end{equation}
where $0 < \rho < 1$ and $P \succ 0$. Note that $\rho, P$ satisfying the first inequality always exist due to the stability of $\mathscr{A}_1$, and given specific $\rho, P$, the the second inequality holds for sufficiently large \( t \) since $\lim_{t\to\infty}( \mathscr{A}_0^t)^\top P \mathscr{A}_0^t = 0$ by the stability of $\mathscr{A}_0$.

Consider a transformed system where the $t$ consecutive steps of applying the fallback control input are combined into one step: denote $\tilde{\mathscr{X}}(j) = \mathscr{X}(i(j))$, where
\begin{equation}
    i(0) = 0, i(j+1) = \begin{cases}
        i(j) + 1 & u(i(j)) = u_1(i(j)), \\
        i(j) + t & \text{otherwise}.
    \end{cases}
\end{equation}
It follows that
\begin{equation}
    \tilde{\mathscr{X}}(j+1) = \tilde{\mathscr{A}}(j)\tilde{\mathscr{X}}(j) + \tilde{\mathscr{W}}(j),
    \label{eq:dyn_tilde}
\end{equation}
where $\tilde{\mathscr{A}}(j), \tilde{\mathscr{W}}(j)$ are defined as:
\begin{equation}
    \tilde{\mathscr{A}}(j) = \begin{cases}
        \mathscr{A}_1 & u(i(j)) = u_1(i(j)), \\
        \mathscr{A}_0^t & \text{otherwise},
    \end{cases}
    \label{eq:At}
\end{equation}
\begin{equation}
    \tilde{\mathscr{W}}(j) = \begin{cases}
        \mathscr{W}(i(j)) \mkern90mu u(i(j)) = u_1(i(j)), \\
        \sum_{\tau = 1}^{t} \mathscr{A}_0^{t - \tau}  \mathscr{W}(i(j) + \tau - 1) \quad \text{otherwise}.
    \end{cases}
    \label{eq:Wt}
\end{equation}

Under the above definitions, we can define the following common Lyapunov function motivated at the beginning of this subsection:
    \begin{equation}
    \tilde{V}(j) = \tilde{\mathscr{X}}_j^\top P \tilde{\mathscr{X}}_j.
\end{equation}
Using the above defined Lyapunov function, Lemma~\ref{lem:moment}, Lemma~\ref{lem:escape} and Theorem~\ref{thm:prop} bound the fourth moment of the state as well as the probability of switching:

\begin{lem}
    It holds for any $j$ that
    \begin{align}
        & \mathbb{E}\tilde{V}(j)^2 \leq \mathcal{Q} := \nonumber \\
        & \quad \frac{6\rho \left( \tr  ( \tilde{\Sigma} P ) \right)^2 + (1 - \rho)(N^2 + 2N) \| P \|^2 \| \tilde{\Sigma} \|^2}{(1 - \rho)(1-\rho^2)},
        \label{eq:Quad}
    \end{align}
    where
    \begin{equation}
        \tilde{\Sigma} := \sum_{\tau = 0}^\infty \mathscr{A}_0^\tau \Sigma \left(\mathscr{A}_0^\tau\right)^\top,
        N = n + n_0 + n_1,
        \label{eq:N}
    \end{equation}
    and $P, \rho$ are defined in~\eqref{eq:common_lyap}.
    \label{lem:moment}
\end{lem}

\begin{lem}
    For any
    \begin{equation}
        a > a_0 := \frac{8N \| \tilde{\Sigma} \| \| P \| \| P^{-1} \|}{1 - \rho^{1/4}},
        \label{eq:a0}
    \end{equation}
    it holds for any $j$ that
    \begin{align}
        & \mathbb{P}\left( \left\| \tilde{ \mathscr{X}}(j) \right\| \geq a \right) \leq \mathcal{E}(a) := \nonumber \\
        & \quad \frac{4N}{\rho^{-1/2}-1}\exp \left( -\frac{(1 - \rho^{1/4})^2}{2N \| \tilde{\Sigma} \| \| P \| \| P^{-1} \|} a^2 \right),
        \label{eq:Escape}
    \end{align}
    where $P, \rho$ are defined in~\eqref{eq:common_lyap}, and $\tilde{\Sigma}$ and $N$ are defined in
    \eqref{eq:N}.
    \label{lem:escape}
\end{lem}

\begin{thm}
    The following properties hold:
    \begin{enumerate}
        \item The fourth moment of the state of the augmented system is bounded:
        \begin{equation}
            \mathbb{E}\left\| \mathscr{X}(k) \right\|_{P_0}^4 \leq 8 \left(  \mathcal{Q} \| P_0 \|_P^2 + (N^2 + 2N)\| P_0 \|_{\tilde{\Sigma}^{-1}}^2 \right),
        \end{equation}
        where $\mathcal{Q}$ is defined in~\eqref{eq:Quad}, the matrices $P_0, P$ are defined in~\eqref{eq:P0} and~\eqref{eq:common_lyap} respectively, and $\tilde{\Sigma}$ and $N$ are defined in
        \eqref{eq:N}.
        \item Let
        \begin{equation}
            \mathcal{K} = \left\| \begin{bmatrix}
                K_0 && -K_1
            \end{bmatrix} \right\|,
        \end{equation}
        and $a_0, \mathcal{E}$ be defined in~\eqref{eq:a0},~\eqref{eq:Escape} respectively, then when the threshold $M \geq a_0 \mathcal{K}$ is large enough, the probability of not using the primary control input satisfies:
        \begin{equation}
            \mathbb{P} \left( u(k) \neq u_1(k) \right) \leq t \mathcal{E}(M / \mathcal{K}),
            \label{eq:Kdiff}
        \end{equation}
        which decays \emph{super-exponentially} w.r.t. the threshold $M$.
    \end{enumerate}
    \label{thm:prop}
\end{thm}

We are now ready to state the main theorem of this subsection:

\begin{thm}
    Let $J_1$ be the LQ cost of the primary controller.
    Assuming that the dwell time $t$ satisfies~\eqref{eq:common_lyap} and that the threshold $M \geq a_0 \mathcal{K}$ is large enough, it holds
    \begin{equation}
        J - J_1 \leq 2c_1c_2 \mathcal{G} + \left(c_2^2 + \| \Delta \|_{P_0} \right) \mathcal{G}^2, 
        \label{eq:gap}
    \end{equation}
    where
    \begin{align}
        & \mathcal{G} = 2^{3/4} \left(  \mathcal{Q} \| P_0 \|_P^2 + (N^2 + 2N)\| P_0 \|_{\tilde{\Sigma}^{-1}}^2 \right)^{1/4} \cdot \nonumber \\
        & \quad (t \mathcal{E}(M / \mathcal{K}))^{1/4}, \\
        & c_1 = \left\| \mathscr{Q}_1 \right\|_P \sqrt{\tr(\Sigma P) / (1 - \rho)}, \\
        & c_2 = \left\|  \mathscr{A}_0 - \mathscr{A}_1 \right\|_{\mathscr{Q}_1+I} \left\| \mathscr{Q}_1+I \right\|_{P_0} \sum_{s=0}^{\infty}  \left\| \mathscr{A}_1^{s} \right\|_{\mathscr{Q}_1+I}, \\
        & \mathscr{Q}_1 = \begin{bmatrix}
            Q & 0 & 0 \\
            0 & 0 & 0 \\
            0 & 0 & K_1^\top R K_1
        \end{bmatrix}, 
        \Delta = \begin{bmatrix}
            0 & 0 & 0 \\
            0 & K_0^\top R K_0 & 0 \\
            0 & 0 & -K_1^\top R K_1
        \end{bmatrix},
    \end{align}
    and the symbols $a_0, \mathcal{K}, P_0, P, \mathcal{Q}, N, \tilde{\Sigma}, \mathcal{E}, \Sigma, \rho,  \mathscr{A}_0, \mathscr{A}_1$
    are defined the same as before in this subsection.
    \label{thm:efficiency}
\end{thm}

The following corollary states that under proper choice of dwell time $t$, the performance loss caused by switching can decay super-exponentially as the switching threshold $M$ is enlarged:

\begin{cor}
    When the primary controller~\eqref{eq:primary} is held constant, and $M, t$ are varied, it holds
    \begin{equation}
        J - J_1 = \mathcal{O}(t^{1/4} \exp(-cM^2))
    \end{equation}
    as $M\to\infty, t\to\infty, t^{1/4} \exp(-cM^2) \to 0$, where $c = (1 - \rho^{1/4})^2 / (2N \| \tilde{\Sigma} \| \| P \| \| P^{-1} \| \mathcal{K}^2)$ is a system-dependent constant.
    \label{col:superexponential}
\end{cor}

\begin{rem}
    The efficacy of the controller parameters \( M \) and \( t \) needed to ensure the efficiency guarantee (as stated in Theorem~\ref{thm:efficiency}) rely on the parameters of the system. When the system parameters are not accessible, it may be difficult to determine the precise minimum values of \( M \) and \( t \); however, a suitable set of parameters can typically be obtained through trial-and-error. Theorem~\ref{thm:safety} guarantees the safety of the closed-loop system for any \((M,t)\), indicating that there is no need to be concerned about destabilizing the system while exploring various values of \((M,t)\). As a result, multiple sets of \((M,t)\) can be attempted until the desired empirical performance is achieved.
\end{rem}

%% file: simulation.tex
\section{Numerical Simulation}
\label{sec:simulation}

In this section, the safety and efficiency of the proposed switching control strategy is verified by simulation on the Tennessee Eastman Process (TEP)~\citep{downs1993plant}, a classical process control system.
In particular, we consider a simplified version of TEP, also used in~\cite{liu2020online}, which has state dimension $n = 8$, input dimension $m = 4$ and output dimension $p = 10$. The LQ weight matrices are $Q = I_n, R = I_m$, and the process and measurement noise distributions are $w_k \sim \mathcal{N}(0, I_n), v_k \sim \mathcal{N}(0, I_p)$. The system is open-loop system, and therefore the fallback controller is chosen as $u_0(k) \equiv 0$.

\subsection{Destabilizing primary controller}

\begin{figure}[!htbp]
    \begin{subfigure}{\columnwidth}
        \centering
        \includegraphics{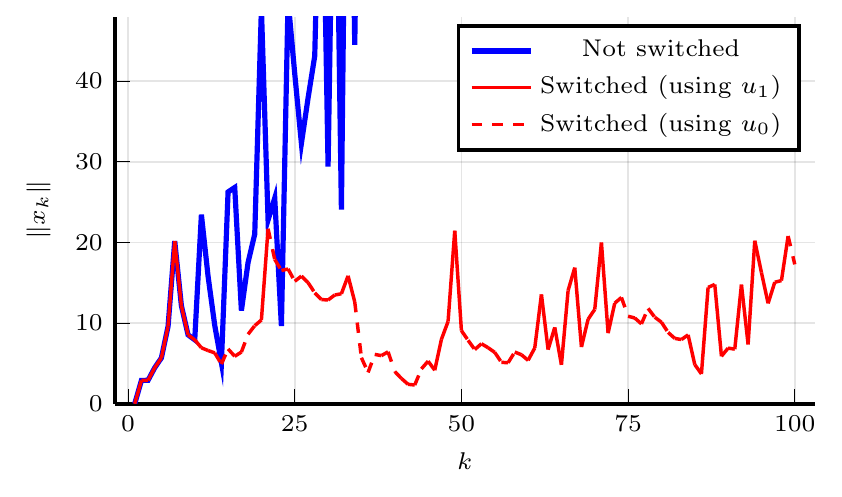}
        \caption{Destabilizing primary controller}
        \label{fig:unstable_traj_compare}
    \end{subfigure}
    \begin{subfigure}{\columnwidth}
        \centering
        \includegraphics{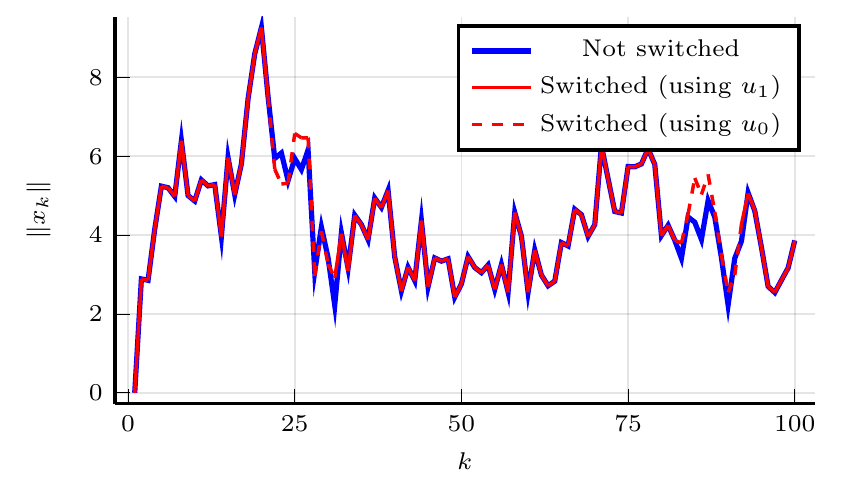}
        \caption{Stabilizing primary controller}
        \label{fig:stable_traj_compare}
    \end{subfigure}
    \caption{Comparison of trajectories of state norms with and without switching, under the same realization of process noise. Parameters of the switching strategy are set to be $M = 1, t = 10$.}
\end{figure}

In this subsection, the primary controller is chosen as $(A_1, B_1, L_1, K_1) = (A-L^*C + \lambda \mathbf{1}_n \mathbf{1}_n^\top, B + \lambda \mathbf{1}_n \mathbf{1}_m^\top, L+\lambda \mathbf{1}_n \mathbf{1}_p^\top, K^* + \lambda \mathbf{1}_m \mathbf{1}_n^\top)$, where $K^*, L^*$ are the optimal feedback gain and Kalman gain respectively, and $\lambda = 0.05$, such that the matrix $\mathscr{A}_1$ is marginally unstable. The trajectories of state norms with and without switching are compared in Fig.~\ref{fig:unstable_traj_compare}. It can be observed that switching effectively prevents the state from growing unboundedly, which qualitatively verifies Theorem~\ref{thm:safety}.

\subsection{Stabilizing primary controller}

\begin{figure}[!htbp]
    \centering
    \includegraphics{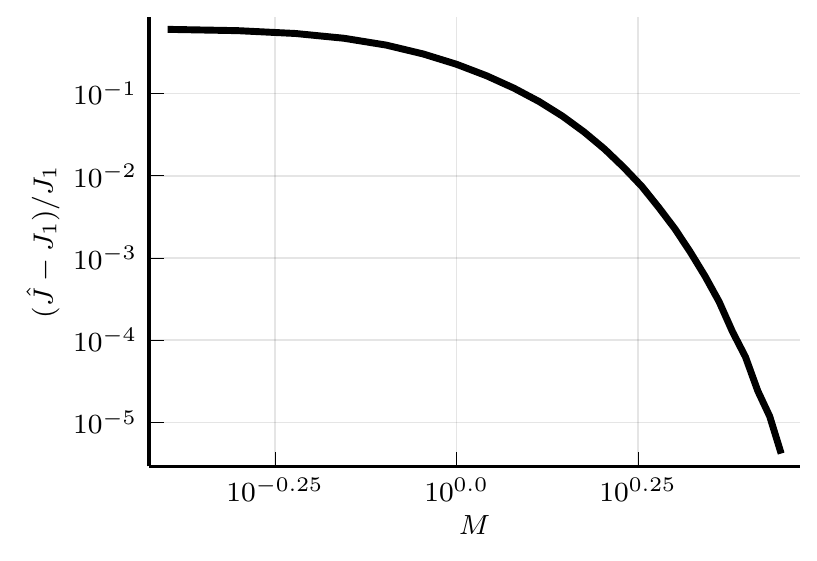}
    \caption{Double-log plot of relative performance gap against switching threshold $M$: super-exponential convergence to zero.}
    \label{fig:gap_wrt_M}
\end{figure}

In this subsection, the primary controller is chosen to be the optimal controller, i.e., $(A_1, B_1, L_1, K_1) = (A - L^*C, B, L^*, K^*)$, where $K^*, L^*$ are the optimal feedback gain and Kalman gain respectively. The state norms with and without switching are compared in Fig.~\ref{fig:stable_traj_compare}, from which it can be observed that switching is activated only occasionally and has a very small effect on the trajectory.
To quantify the performance loss caused by switching and its relationship with the threshold $M$, we fix $t = 10$ and increase $M$ from $0.5$ to $3$. We evaluate the relative performance loss $(J - J_1) / J_1$ for each $M$ using the empirical average of $10^5$ trajectories, each of length $10^3$, and plot the relationship in the double-log graph shown in Fig.~\ref{fig:gap_wrt_M}. It can be observed that the performance loss decays to zero faster than exponential convergence (i.e., a straight line in the double-log plot), which verifies Theorem~\ref{thm:efficiency} and Corollary~\ref{col:superexponential}.

%% file: conclusion.tex
\section{Conclusion}
\label{sec:conclusion}

This manuscript proposes and analyzes a control strategy for partially observed linear-Gaussian systems which switches between an uncertified primary controller and a stabilizing albeit conservative fallback controller in pursuit of both safety and efficiency.
It is guaranteed that the LQ cost is bounded regardless of how the primary controller is chosen.
Furthermore, the extra cost caused by switching is quantified as decaying super-exponentially to zero as the threshold for triggering the switching increases.
An interesting future research topic would be to apply the proposed switching strategy as a ``plug-and-play'' modification to existing adaptive LQG algorithms for end-to-end stability and performance guarantees.

%% file: appendix.tex
\section{Proof of Lemma~\ref{lem:V0}}

\begin{pf}
    From the switching strategy, it holds
    \begin{equation}
        \begin{bmatrix}
            x(k+1) \\
            z_0(k+1)
        \end{bmatrix} = \mathcal{A}_0
        \begin{bmatrix}
            x(k) \\
            z_0(k)
        \end{bmatrix} + \begin{bmatrix}
            w(k) \\
            L_0 v(k)
        \end{bmatrix} + d(k),
    \end{equation}
    where $d(k) = \mathbf{1}_{\{ u(k) = u_1(k) \}} [B^\top B_0^\top]^\top (u_1(k) - u_0(k))$ satisfies $\| d(k) \| \leq M \| [B^\top B_0^\top] \|$, and hence
    \begin{equation}
        \| d(k) \|_{P_0} \leq M \| [B^\top B_0^\top] \| \| P_0 \|^{1/2}.
    \end{equation}
    Therefore, it holds
    \begin{align}
        & V_0(k+1)  =  \left\| \begin{bmatrix}
            x(k+1) \\
            z_0(k+1)
        \end{bmatrix} \right\|_{P_0}^2 \nonumber \\
        & \leq  \left( \left\| \mathcal{A}_0
        \begin{bmatrix}
            x(k) \\
            z_0(k)
        \end{bmatrix} \right\|_{P_0} + \left\| d(k) + \begin{bmatrix}
            w(k) \\
            L_0 v(k)
        \end{bmatrix} \right\|_{P_0} \right)^2 \nonumber \\
        & = (1+\sigma) \left\| \mathcal{A}_0
        \begin{bmatrix}
            x(k) \\
            z_0(k)
        \end{bmatrix} \right\|_{P_0}^2 +  \nonumber \\
        & \quad \left( 1+\frac{1}{\sigma} \right)\left\| d(k) + \begin{bmatrix}
            w(k) \\
            L_0 v(k)
        \end{bmatrix} \right\|_{P_0}^2 \nonumber \\
        & \leq (1+\sigma) \rho_0 V_0(k) +  \left( 1+\frac{1}{\sigma} \right)\left\| d(k) + \begin{bmatrix}
            w(k) \\
            L_0 v(k)
        \end{bmatrix} \right\|_{P_0}^2,
        \label{eq:Vk1}
    \end{align}
    where $\sigma = (\rho_0^{-1} - 1)/2$, and the last inequality follows from~\eqref{eq:rho0}.
    Notice that
    \begin{align}
        & \mathbb{E}\left\| d(k) + \begin{bmatrix}
            w(k) \\
            L_0 v(k)
        \end{bmatrix} \right\|_{P_0}^2 \leq 2\mathbb{E} \left( \| d_k \|_{P_0}^2 + \left\| \begin{bmatrix}
            w(k) \\
            L_0 v(k)
        \end{bmatrix} \right\|_{P_0}^2 \right) \nonumber \\
        & \leq 2 \left( \mathbb{E}\| d_k \|_{P_0}^2 + \mathbb{E}\left\| \begin{bmatrix}
            w(k) \\
            L_0 v(k)
        \end{bmatrix} \right\|_{P_0}^2 \right) \nonumber \\
        & \leq 2 \left(  M^2 \left\| \begin{bmatrix} B \\ B_0 \end{bmatrix} \right\|^2 \| P_0 \| + \tr \left( \begin{bmatrix}
            W & 0 \\
            0 & L_0 VL_0^\top
        \end{bmatrix} P_0 \right) \right).
    \end{align}
    Therefore, it follows from~\eqref{eq:Vk1} and the induction on $k$ that
    \begin{align}
        & \mathbb{E}V_0(k) \leq
        \frac{2\left( 1 + \frac{1}{\sigma} \right)}{1 - (1+\sigma)\rho_0} \cdot\nonumber  \\
        & \left(  M^2 \left\|\begin{bmatrix} B \\ B_0 \end{bmatrix} \right\|^2 \| P_0 \| + \tr \left( \begin{bmatrix}
            W & 0 \\
            0 & L_0VL_0^\top
        \end{bmatrix} P_0 \right) \right)\nonumber  \\
        &=
        \frac{4(1+\rho_0)}{(1 - \rho_0)^2}\cdot \nonumber  \\& \left(  M^2 \left\|\begin{bmatrix} B \\ B_0 \end{bmatrix} \right\|^2 \| P_0 \| + \tr \left( \begin{bmatrix}
            W & 0 \\
            0 & L_0VL_0^\top
        \end{bmatrix} P_0 \right) \right).
    \end{align}
    Therefore, the conclusion follows from~\eqref{eq:Vk1} and the induction on $k$.
\end{pf}

\section{Proof of Theorem~\ref{thm:safety}}

\begin{pf}
    By definition of $J$, we only need to prove $\mathbb{E}(\| x(k) \|_Q^2 + \| u(k) \|_R^2)$ is not greater than the RHS of~\eqref{eq:safe} for any $k$.
    Notice that
    \begin{align}
        & \| x(k) \|_Q^2 + \| u(k) \|_R^2 \nonumber \\
        & \leq \left\| \begin{bmatrix}
            x(k) \\ z_0(k)
        \end{bmatrix} \right\|_{
            \begin{bmatrix}
                Q & 0 \\
                0 & K_0^\top R K_0 + I
            \end{bmatrix}
        }^2 + \| u(k) \|_R^2 - \| K_0 z_0(k) \|_R^2 \nonumber \\
        & \leq \left\| \begin{bmatrix}
            x(k) \\ z_0(k)
        \end{bmatrix} \right\|_{P_0}^2 + \| u(k) \|_R^2 - \| u_0(k) \|_R^2.
        \label{eq:xu}
    \end{align}
    From the switching strategy, it holds $\| u(k) - u_0(k) \| \leq M$, and hence,
    \begin{equation}
        \| u(k) \|_R \leq \| u_0(k) \|_R + M \| R \|^{1/2},
    \end{equation}
    which implies
    \begin{align}
        & \| u(k) \|_R^2 - \| u_0(k) \|_R^2 \leq 2M \| R \|^{1/2} \| u_0(k) \|_R + M^2 \| R \| \nonumber \\
        & \leq 2M \| R \|^{1/2}  \left\| \begin{bmatrix}
            x(k) \\ z_0(k)
        \end{bmatrix} \right\|_{P_0} + M^2 \| R \|.
        \label{eq:udiff}
    \end{align}
    Substituting~\eqref{eq:udiff} into~\eqref{eq:xu}, we get
    \begin{align}
        & \| x(k) \|_Q^2 + \| u(k) \|_R^2 \leq  \left( \left\| \begin{bmatrix}
            x(k) \\ z_0(k)
        \end{bmatrix} \right\|_{P_0} + M \| R \|^{1/2} \right)^2\nonumber  \\
        & \leq 2(V_0(k) + M^2 \| R \|).
    \end{align}
    The conclusion then follows from Lemma~\ref{lem:V0}.
\end{pf}

\section{Proof of Lemma~\ref{lem:moment}}

\begin{pf}
    From~\eqref{eq:common_lyap} and~\eqref{eq:dyn_tilde}, it follows that
    \begin{equation}
        \tilde{V}(j+1) \leq \rho \tilde{V}(j) + \eta(j),
        \label{eq:V_rec}
    \end{equation}
    where
    \begin{equation}
        \eta(j) = 2 \tilde{\mathscr{W}}(j) ^\top P \tilde{\mathscr{A}}(j) \tilde{\mathscr{X}}(j) + \tilde{\mathscr{W}}(j) ^\top P\tilde{\mathscr{W}}(j).
    \end{equation}
    From
    \begin{align}
        & \mathbb{E}\left[ \tilde{\mathscr{W}}(j) ^\top P \tilde{\mathscr{A}}(j) \tilde{\mathscr{X}}(j) \right] =
        \nonumber \\
        & \quad \mathbb{E} \left[ \mathbb{E}\left[ \tilde{\mathscr{W}}(j) \mid \tilde{\mathscr{X}}(j) \right]^\top P \tilde{\mathscr{A}}(j) \tilde{\mathscr{X}}(j) \right] =
        0,
    \end{align}
    it follows that
    \begin{equation}
        \mathbb{E}\eta(j) = \tr \left( \mathbb{E} \left[ \tilde{\mathscr{W}}(j)\tilde{\mathscr{W}}(j)^\top \right]P \right) \leq \tr(\tilde\Sigma P),
    \end{equation}
    and hence,
    \begin{equation}
        \mathbb{E}\tilde{V}(j) \leq \tr(\tilde\Sigma P) / (1 - \rho).
        \label{eq:EVt}
    \end{equation}
    Now squaring both sides of~\eqref{eq:V_rec} and taking the expectations, we obtain
    \begin{equation}
        \mathbb{E}\tilde{V}(j+1)^2 \leq \rho^2 \mathbb{E}\tilde{V}(j)^2 + 2\rho \mathbb{E} \left[ \tilde{V}(j)\eta(j) \right] + \mathbb{E}\eta(j)^2.
        \label{eq:EVt2_rec}
    \end{equation}

    \begin{enumerate}
        \item Bound on $\mathbb{E} \left[ \tilde{V}(j)\eta(j) \right]$:
        \begin{align}
            & \mathbb{E} \left[ \tilde{V}(j)\eta(j) \right] = 
            2 \mathbb{E}  \left[ \tilde{\mathscr{W}}(j) ^\top P \tilde{\mathscr{A}}(j) \tilde{\mathscr{X}}(j) \tilde{V}(j)\right] +
            \nonumber \\
            & \quad \mathbb{E} \left[ \tilde{\mathscr{W}}(j) ^\top P\tilde{\mathscr{W}}(j)\tilde{V}(j) \right] \nonumber \\
            & = 
            2\mathbb{E}  \left[  \underbrace{\mathbb{E}\left[\tilde{\mathscr{W}}(j) \mid \tilde{\mathscr{X}}(j)\right] ^\top}_{=0} P \tilde{\mathscr{A}}(j) \tilde{\mathscr{X}}(j) \tilde{V}(j)\right] + \nonumber \\
            & \quad \tr \left(  \mathbb{E} \tilde{V}(j)  \mathbb{E}\left[ \tilde{\mathscr{W}}(j)\tilde{\mathscr{W}}(j)^\top \right]P\right) \nonumber \\
            & = \tr(\tilde\Sigma P) \mathbb{E}\tilde{V}(j).
            \label{eq:EVeta}
        \end{align}

        \item Bound on $\mathbb{E}\eta(j)^2$:
        \begin{align}
            & \mathbb{E}\eta(j)^2
            = \nonumber \\
            & \quad 4 \mathbb{E}\left[\tilde{\mathscr{X}}(j)^\top \tilde{\mathscr{A}}(j)^\top P \tilde{\mathscr{W}}(j)\tilde{\mathscr{W}}(j)^\top P \tilde{\mathscr{A}}(j) \tilde{\mathscr{X}}(j)\right] + \nonumber  \\
            & \quad 4  \mathbb{E} \left[ \tilde{\mathscr{W}}(j)^\top P \tilde{\mathscr{A}}(j) \tilde{\mathscr{X}}(j) \tilde{\mathscr{W}}(j)^\top P \tilde{\mathscr{W}}(j) \right] + \nonumber \\
            & \quad \mathbb{E} \left[ \tilde{\mathscr{W}}(j)^\top P \tilde{\mathscr{W}}(j)\tilde{\mathscr{W}}(j)^\top P \tilde{\mathscr{W}}(j) \right] \nonumber \\
            &
            = 4 \tr \left( \mathbb{E}\left[ \tilde{\mathscr{W}}(j)\tilde{\mathscr{W}}(j)^\top \right]P \right) \cdot \nonumber \\
            & \qquad \mathbb{E}\left[\tilde{\mathscr{X}}(j)^\top \tilde{\mathscr{A}}(j)^\top  P \tilde{\mathscr{A}}(j) \tilde{\mathscr{X}}(j)\right] + \nonumber \\
            & \quad 4\tr \left( \mathbb{E} \left[ \tilde{\mathscr{A}}(j) \tilde{\mathscr{X}}(j) \cdot \vphantom{\underbrace{\mathbb{E} \left[ \tilde{\mathscr{W}}(j)^\top P\tilde{\mathscr{W}}(j)\tilde{\mathscr{W}}(j)^\top \mid \tilde{\mathscr{A}}(j) \right]}_{=0 \text{ by symmetry}}} \right.\right. \nonumber \\
            & \qquad \left.\left. \underbrace{\mathbb{E} \left[ \tilde{\mathscr{W}}(j)^\top P\tilde{\mathscr{W}}(j)\tilde{\mathscr{W}}(j)^\top \mid \tilde{\mathscr{A}}(j) \right]}_{=0 \text{ by symmetry}} \right] \right) + \nonumber \\
            & \quad \| P \|^2 \| \tilde\Sigma \|^2 \underbrace{\mathbb{E} \nu^2}_{\text{where } \nu \sim \chi^2(N)} \nonumber \\
            & \leq 4\rho \tr(\tilde\Sigma P) \mathbb{E}\tilde{V}(j) + (N^2+2N) \| P \|^2 \| \tilde\Sigma \|^2.
            \label{eq:Eeta2}
        \end{align}
    \end{enumerate}
    Combining~\eqref{eq:EVt}, \eqref{eq:EVt2_rec}, \eqref{eq:EVeta}, \eqref{eq:Eeta2} and applying induction leads to the conclusion.
\end{pf}

\section{Proof of Lemma~\ref{lem:escape}}

\begin{pf}
    Notice that
    \begin{equation}
        \tilde{ \mathscr{X}}(j) = \sum_{s=0}^{j-1} \left( \prod_{r=s+1}^{k-1} \tilde{\mathscr{A}}(r) \right) \tilde{\mathscr{W}}(s).
    \end{equation}
    From~\eqref{eq:common_lyap} and~\eqref{eq:At}, it follows that
    \begin{align}
        & \left\| \tilde{ \mathscr{X}}(j) \right\|_P = \left\| \sum_{s=0}^{j-1} P^{1/2} \left( \prod_{r=s+1}^{k-1} \tilde{\mathscr{A}}(r) \right) \tilde{\mathscr{W}}(s)\right\| \nonumber \\
        & \leq \left\| \sum_{s=0}^{j-1} \rho^{(j - s - 1)/2} P^{1/2}  \tilde{\mathscr{W}}(s)\right\|  \leq \sum_{s=0}^{j-1} \rho^{(j - s - 1)/2} \left\|  \tilde{\mathscr{W}}(s)\right\|_P.
    \end{align}
    By~\eqref{eq:Wt}, it holds $\tilde{\mathscr{W}}(s) \mid \mathcal{F}(s - 1) \sim \mathcal{N}(0, \Sigma(s))$, where $\mathcal{F}(s - 1)$ is the $\sigma$-algebra generated by $\tilde{\mathscr{W}}(0), \ldots, \tilde{\mathscr{W}}(s - 1)$, and
    \begin{equation}
        \Sigma(s) \in \left\{ \Sigma, \sum_{\tau = 0}^{t - 1} \mathscr{A}_0^\tau \Sigma \left(\mathscr{A}_0^\tau\right)^\top \right\};
    \end{equation}
    in either case, it holds $\Sigma(s) \preceq \tilde{\Sigma}$.
    Hence, by a concentration bound on Gaussian random vectors~\cite[Lemma 3.1]{ledoux1991probability}, it holds for any $s$ and any $a > 0$ that
    \begin{equation}
        \mathbb{P} \left( \left\|\tilde{\mathscr{W}}(s)\right\|_P \geq a \right) \leq 2n \exp \left( -\frac{(1 - \rho^{1/4})^2}{2N \| \tilde{\Sigma} \| \| P \|} a^2 \right).
    \end{equation}
    Invoking a tail bound on the exponentially weighted sum of Gaussian-like random variables~\cite[Theorem 3]{cdc} with $\varrho = \rho^{1/2}$, and assuming w.l.o.g. that $\rho \in (1/4, 1)$, it follows that
    \begin{equation}
        \mathbb{P} \left( \left\| \tilde{ \mathscr{X}}(j) \right\|_P \geq a \right) \leq \frac{4N}{\rho^{-1/2} - 1} \exp \left( -\frac{(1 - \rho^{1/4})^2}{2N \| \tilde{\Sigma} \| \| P \|} a^2 \right).
    \end{equation}
    Meanwhile, it holds
    \begin{equation}
         \left\{ \left\|\tilde{ \mathscr{X}}(j)\right\| \geq a \right\} \subseteq \left\{ \left\|\tilde{ \mathscr{X}}(j)\right\|_P \geq a \| P^{-1} \|^{1/2}\right\},
    \end{equation}
    from which the conclusion follows.
\end{pf}

\section{Proof of Theorem~\ref{thm:prop}}

\begin{pf}
    The proof is devoted to translating properties of the transformed system $\{ \tilde{ \mathscr{X}}(j) \}$ (Lemma~\ref{lem:moment} and Lemma~\ref{lem:escape}) back into the properties of the original system.
    \begin{enumerate}
        \item Let $j = \sup \{ s \in \mathbb{N} \mid i(s) \leq k \}$, i.e., $\tilde{ \mathscr{X}}(j)$ is the last state in the transformed state sequence that occurs no later than $\mathscr{X}(k)$.
        Consequently,
        \begin{equation}
            \mathscr{X}(k) = \mathscr{A}_0^{k - i(j)} \tilde{ \mathscr{X}}(j) + \tilde{\mathscr{W}}_{jk},
        \end{equation}
        where $\tilde{\mathscr{W}}_{jk}$ is defined as:
        \begin{equation}
            \tilde{\mathscr{W}}_{jk} := \sum_{\tau=1}^{k - i(j)}\mathscr{A}_0^{k - i(j) - \tau} \mathscr{W}(i(j) + \tau - 1).
        \end{equation}
        From~\eqref{eq:P0}, it follows that
        \begin{align}
             \left\| \mathscr{X}(k) \right\|_{P_0} & \leq \rho_0^{(k - i(j)) / 2} \left\| \tilde{ \mathscr{X}}(j) \right\|_{P_0} + \left\| \tilde{\mathscr{W}}_{jk} \right\|_{P_0} \nonumber \\
             & \leq \left\| \tilde{ \mathscr{X}}(j) \right\|_{P_0} + \left\| \tilde{\mathscr{W}}_{jk} \right\|_{P_0}.
             \label{eq:xP0}
        \end{align}
        Hence, by applying the power means inequality $((a+b)/2)^4 \leq (a^4 + b^4) / 2$, and taking the expectations on both sides of~\eqref{eq:xP0}, we have
        \begin{equation}
            \mathbb{E}\left\| \mathscr{X}(k) \right\|_{P_0}^4 \leq 8 \left(  \mathbb{E}\left\| \tilde{ \mathscr{X}}(j) \right\|_{P_0}^4 + \mathbb{E}\left\| \tilde{\mathscr{W}}_{jk} \right\|_{P_0}^4 \right).
            \label{eq:ExP04}
        \end{equation}
        The terms in the RHS of~\eqref{eq:ExP04} can be bounded as follows:
        \begin{itemize}
            \item $\mathbb{E}\left\| \tilde{ \mathscr{X}}(j) \right\|_{P_0}^4 \leq \mathcal{Q} \| P_0 \|_P^2$, which follows from Lemma~\ref{lem:moment}.
            \item $\mathbb{E}\left\| \tilde{\mathscr{W}}_{jk} \right\|_{P_0}^4 \leq \| P_0 \|_{\tilde{\Sigma}^{-1}}^2 \mathbb{E} \left\| \tilde{\mathscr{W}}_{jk} \right\|_{\tilde{\Sigma}^{-1}}^4 \leq \| P_0 \|_{\tilde{\Sigma}^{-1}}^2 \cdot \mathbb{E}\nu^2  = (N^2 + 2N)\| P_0 \|_{\tilde{\Sigma}^{-1}}^2$, where $\nu \sim \chi^2(N)$, since $\tilde{\mathscr{W}}_{jk}$ is Gaussian distributed with zero mean and covariance no greater than $\tilde{\Sigma}$.
        \end{itemize}
        Combining the above two items leads to the conclusion.

        \item Define the index set
        \begin{equation}
            \mathcal{I} = \left\{ k \in \mathbb{N} \mid \exists j \in \mathbb{N}, \text{ s.t. } i(j) = k \right\},
        \end{equation}
        which are the indices of states $\mathscr{X}(k)$ that occur in the transformed state sequence $\{ \tilde{ \mathscr{X}}(j) \}$.
        Since a sufficient and necessary condition of $u(k) \neq u_1(k)$ is that exactly one of $\mathscr{X}(k), \mathscr{X}(k-1), \ldots, \mathscr{X}(k - t+1)$ belongs to the transformed state sequence and triggers the switching rule, it holds
        \begin{align}
            & \left\{ u(k) \neq u_1(k) \right\} \subseteq \bigcup_{\tau = 0}^{t-1} \nonumber \\
            & \quad \left\{ \| u_0(k - \tau) - u_1(k - \tau) \| \geq M, k - \tau \in \mathcal{I} \right\}.
            \label{eq:unequ1}
        \end{align}
        For each event in the RHS of~\eqref{eq:unequ1}, we have
        \begin{align}
            & \mathbb{P} \left( \| u_0(k - \tau) - u_1(k - \tau) \| \geq M, k - \tau \in \mathcal{I} \right) \nonumber \\
            &= \mathbb{P} \left( \| u_0(k - \tau) - u_1(k - \tau) \| \geq M| k - \tau \in \mathcal{I} \right) \nonumber \\
            & \quad \mathbb{P}(k - \tau \in \mathcal{I}) \nonumber \\
            &\leq \mathbb{P} \left( \| u_0(k - \tau) - u_1(k - \tau) \| \geq M| k - \tau \in \mathcal{I} \right),
        \end{align}
        and since $u_0(k) - u_1(k) = \begin{bmatrix}
            0 & K_0 & -K_1
        \end{bmatrix} \mathscr{X}(k)$ for any $k$, we have $\| u_0(k) - u_1(k) \| \leq \mathcal{K} \| \mathscr{X}(k) \|$ for any $k$, which further implies
        \begin{align}
            & \mathbb{P} \left( \| u_0(k - \tau) - u_1(k - \tau) \| \geq M, k - \tau \in \mathcal{I} \right) \nonumber \\
            & \leq  \mathbb{P} \left( \| \mathscr{X}(k - \tau) \| \geq M / \mathcal{K} \mid k - \tau \in \mathcal{I} \right).
            \label{eq:escape_udiff}
        \end{align}
        Since $\mathbb{P}\left( \left\| \tilde{\mathscr{X}}(j) \right\| \geq M / \mathcal{K} \right) \leq \mathcal{E}(M / \mathcal{K})$ for any $j$ according to Lemma~\ref{lem:escape}, and $k - \tau \in \mathcal{I}$ indicates that $\mathscr{X}(k - \tau)$ belongs to $\left\{ \tilde{\mathscr{X}}(j) \right\}$, it follows that the RHS of~\eqref{eq:escape_udiff} is not greater than $\mathcal{E}(M / \mathcal{K})$. The conclusion then follows from~\eqref{eq:unequ1} by taking the union bound over $\tau = 0,1,\ldots, t-1$.
    \end{enumerate}
\end{pf}

\section{Proof of Theorem~\ref{thm:efficiency}}

\begin{pf}
    Let $\check{\mathscr{X}}(0) =  \mathscr{X}(0)$ and $\check{\mathscr{X}}(k+1) = \mathscr{A}_1 \check{\mathscr{X}}(k) + \mathscr{W}(k)$, i.e., the sequence $\left\{ \check{\mathscr{X}}(k) \right\}$ is the state sequence if switching is not applied, then
    \begin{equation}
        J_1 = \lim_{T\to\infty} \frac{1}{T} \sum_{k = 0}^{T - 1} \mathbb{E} \left\|  \mathscr{Q}_1^{1/2} \check{\mathscr{X}}(k) \right\|^2.
    \end{equation}
    On the other hand, we have
    \begin{equation}
        J = \limsup_{T \to \infty} \frac{1}{T} \sum_{k=0}^{T-1} \mathbb{E} \left[ \| x(k) \|_Q^2 + \| u(k) \|_R^2 \right].
    \end{equation}
    Therefore, we only need to prove that $$\mathbb{E} \left[ \| x(k) \|_Q^2 + \| u(k) \|_R^2 -  \left\|  \mathscr{Q}_1^{1/2} \check{\mathscr{X}}(k) \right\|^2\right]$$ is no greater than the RHS of~\eqref{eq:gap} for any $k$.
    Notice that
    \begin{align}
        & \| x(k) \|_Q^2 + \| u(k) \|_R^2 -  \left\|  \mathscr{Q}_1^{1/2} \check{\mathscr{X}}(k) \right\|^2
        = \left\|  \mathscr{Q}_1^{1/2} \mathscr{X}(k) \right\|^2-
        \nonumber \\ &\quad 
        \left\|  \mathscr{Q}_1^{1/2} \check{\mathscr{X}}(k) \right\|^2 +
        \left\| \Delta^{1/2} \mathscr{X}(k) \right\|^2 \mathbf{1}_{ \left\{ u(k) \neq u_1(k) \right\}}.
        \label{eq:stage_cost_diff}
    \end{align}
    Next we shall bound $ \mathbb{E} \left[\left\|  \mathscr{Q}_1^{1/2} \mathscr{X}(k) \right\|^2-  \left\|  \mathscr{Q}_1^{1/2} \check{\mathscr{X}}(k) \right\|^2 \right]$ and $ \mathbb{E}\left[\left\| \Delta^{1/2} \mathscr{X}(k) \right\|^2 \mathbf{1}_{ \left\{ u(k) \neq u_1(k) \right\}}\right]$ respectively:

    \begin{enumerate}
        \item Bounding $ \mathbb{E}\left[\left\|  \mathscr{Q}_1^{1/2} \mathscr{X}(k) \right\|^2-  \left\|  \mathscr{Q}_1^{1/2} \check{\mathscr{X}}(k) \right\|^2\right]$: Notice that
        \begin{align}
            & \mathscr{X}(k) = \mathscr{A}_1 \mathscr{X}(k - 1) + \mathscr{W}(k - 1) + \nonumber \\
            & \quad ( \mathscr{A}_0 - \mathscr{A}_1)  \mathscr{X}(k - 1) \mathbf{1}_{ \left\{ u(k) \neq u_1(k) \right\}},
        \end{align}
        and by recursively applying this expansion, we get
        \begin{align}
            & \mathscr{X}(k) = \mathscr{A}_1^k \mathscr{X}(0) + \sum_{s=0}^{k - 1}  \mathscr{A}_1^{k - s - 1} \cdot \nonumber \\
            & \quad  \left( \mathscr{W}(s) + ( \mathscr{A}_0 - \mathscr{A}_1)  \mathscr{X}(s) \mathbf{1}_{ \left\{ u(k) \neq u_1(k) \right\}} \right) \nonumber \\
            & = \check{\mathscr{X}}(k) + \sum_{s=0}^{k - 1}  \mathscr{A}_1^{k - s - 1}( \mathscr{A}_0 - \mathscr{A}_1)  \mathscr{X}(s) \mathbf{1}_{ \left\{ u(k) \neq u_1(k) \right\}}.
        \end{align}
        Hence,
        \begin{align}
            & \left\| \mathscr{Q}_1^{1/2} \mathscr{X}(k) \right\| \leq \left\| \mathscr{Q}_1^{1/2} \check{\mathscr{X}}(k) \right\| + \nonumber \\
            & \quad \left\| \sum_{s=0}^{k - 1}  \mathscr{A}_1^{k - s - 1}( \mathscr{A}_0 - \mathscr{A}_1)  \mathscr{X}(s) \mathbf{1}_{ \left\{ u(k) \neq u_1(k) \right\}} \right\|_{\mathscr{Q}_1+I} \nonumber \\
            &  \leq \left\| \mathscr{Q}_1^{1/2} \check{\mathscr{X}}(k) \right\| + \left\|  \mathscr{A}_0 - \mathscr{A}_1 \right\|_{\mathscr{Q}_1+I} \cdot \nonumber \\
            & \quad \sum_{s=0}^{k-1}  \left\| \mathscr{A}_1^{k - s - 1} \right\|_{\mathscr{Q}_1+I} \| \mathscr{X}(s) \|_{\mathscr{Q}_1+I}\mathbf{1}_{ \left\{ u(k) \neq u_1(k) \right\}}.
        \end{align}
        From the fact that $\mathbb{E}\left(\sum_{i=1}^n X_i\right)^2 \leq\left(\sum_{i=1}^n \sqrt{\mathbb{E} X_i^2}\right)^2$ for any random variables $X_1, \ldots, X_n$, we have
        \begin{align}
             & \mathbb{E}\left\| \mathscr{Q}_1^{1/2} \mathscr{X}(k) \right\| ^2 \leq \left( \sqrt{\mathbb{E}\left\| \mathscr{Q}_1^{1/2} \check{\mathscr{X}}(k) \right\| ^2} + \right. \nonumber \\
             & \quad \left\|  \mathscr{A}_0 - \mathscr{A}_1 \right\|_{\mathscr{Q}_1+I} \left\| \mathscr{Q}_1+I \right\|_{P_0} \sum_{s=0}^{k-1}  \left\| \mathscr{A}_1^{k - s - 1} \right\|_{\mathscr{Q}_1+I} \nonumber \\
             & \quad \left. \vphantom{\sqrt{\mathbb{E}\left\| \mathscr{Q}_1^{1/2} \check{\mathscr{X}}(k) \right\| ^2}} \sqrt{\mathbb{E} \left[  \| \mathscr{X}(s) \|_{P_0}^2 \mathbf{1}_{ \left\{ u(k) \neq u_1(k) \right\}} \right]}\right)^2.
             \label{eq:EQ1X2}
        \end{align}
        By~\eqref{eq:common_lyap}, we have $\mathbb{E} \left\| \check{\mathscr{X}}(k) \right\|_P^2 \leq \tr(\Sigma P) / (1 - \rho)$,
        and hence,
        \begin{equation}
            \mathbb{E} \left\| \mathscr{Q}_1^{1/2} \check{\mathscr{X}}(k) \right\| ^ 2 \leq \frac{\tr(\Sigma P) \left\| \mathscr{Q}_1 \right\|_P^2}{1 - \rho} = c_1^2.
            \label{eq:EQ1Xc2}
        \end{equation}
        Meanwhile, by Cauchy-Schwarz inequality and Theorem~\ref{thm:prop}, it holds
        \begin{align}
            & \mathbb{E} \left[  \| \mathscr{X}(s) \|_{P_0}^2 \mathbf{1}_{ \left\{ u(k) \neq u_1(k) \right\}} \right] \leq \nonumber \\
            & \quad \sqrt{ \mathbb{E}  \| \mathscr{X}(s) \|_{P_0}^4 \mathbb{P}(u(k)\neq u_1(k))} \leq \mathcal{G}^2.
            \label{eq:Ecross}
        \end{align}
        Combining~\eqref{eq:EQ1X2}, \eqref{eq:EQ1Xc2} and~\eqref{eq:Ecross} leads to
        \begin{equation}
            \mathbb{E}\left\| \mathscr{Q}_1^{1/2} \mathscr{X}(k) \right\|^2 - \mathbb{E} \left\| \mathscr{Q}_1^{1/2} \check{\mathscr{X}}(k) \right\| ^ 2 \leq 2c_1c_2 \mathcal{G} + c_2^2 \mathcal{G}^2.
            \label{eq:stage_cost_diff_1}
        \end{equation}

        \item Bounding $ \mathbb{E}\left[\left\| \Delta^{1/2} \mathscr{X}(k) \right\|^2 \mathbf{1}_{ \left\{ u(k) \neq u_1(k) \right\}}\right]$: similarly to~\eqref{eq:Ecross}, by Cauchy-Schwarz inequality and Theorem~\ref{thm:prop}, we have
        \begin{align}
            & \mathbb{E}\left[\left\| \Delta^{1/2} \mathscr{X}(k) \right\|^2 \mathbf{1}_{ \left\{ u(k) \neq u_1(k) \right\}}\right]
            \nonumber \\
            & \leq \left\| \Delta \right\|_{P_0} \mathbb{E} \left[  \| \mathscr{X}(k) \|_{P_0}^2 \mathbf{1}_{ \left\{ u(k) \neq u_1(k) \right\}} \right] \nonumber \\
            & \leq \left\| \Delta \right\|_{P_0}\sqrt{ \mathbb{E}  \| \mathscr{X}(k) \|_{P_0}^4 \mathbb{P}(u(k)\neq u_1(k))}
            \leq \left\| \Delta \right\|_{P_0}\mathcal{G}^2.
            \label{eq:stage_cost_diff_2}
        \end{align}
    \end{enumerate}

    Substituting~\eqref{eq:stage_cost_diff_1} and~\eqref{eq:stage_cost_diff_2} into~\eqref{eq:stage_cost_diff} leads to the conclusion.
\end{pf}